\documentclass[sn-vancouver,smallextended]{sn-jnl}

\jyear{2023}%

\theoremstyle{thmstyleone}%
\theoremstyle{thmstyletwo}%

\theoremstyle{thmstylethree}%

\usepackage{mdframed}

\begin{document}
\title[AI to automate the systematic review of scientific literature]{Artificial intelligence to automate the systematic review of scientific literature\footnote{Published version available at: 
\url{https://doi.org/10.1007/s00607-023-01181-x}
}}


\author[1]{\fnm{Jos\'e} \sur{de la Torre-L\'opez}}\email{i82toloj@uco.es}

\author[1]{\fnm{Aurora} \sur{Ram\'irez}\footnote{Corresponding author}}\email{aramirez@uco.es}

\author[1]{\fnm{Jos\'e Ra\'ul} \sur{Romero}}\email{jrromero@uco.es}

\affil*[1]{\orgdiv{Dept. Computer Science and Numerical Analysis}, \orgname{University of C\'ordoba}, \orgaddress{\street{Rabanales Campus}, \city{C\'ordoba}, \postcode{14071}, \country{Spain}}}


\abstract{Artificial intelligence (AI) has acquired notorious relevance in modern computing as it effectively solves complex tasks traditionally done by humans. AI provides methods to represent and infer knowledge, efficiently manipulate texts and learn from vast amount of data. These characteristics are applicable in many activities that human find laborious or repetitive, as is the case of the analysis of scientific literature. Manually preparing and writing a systematic literature review (SLR) takes considerable time and effort, since it requires planning a strategy, conducting the literature search and analysis, and reporting the findings. Depending on the area under study, the number of papers retrieved can be of hundreds or thousands, meaning that filtering those relevant ones and extracting the key information becomes a costly and error-prone process. However, some of the involved tasks are repetitive and, therefore, subject to automation by means of AI. In this paper, we present a survey of AI techniques proposed in the last 15 years to help researchers conduct systematic analyses of scientific literature. We describe the tasks currently supported, the types of algorithms applied, and available tools proposed in 34 primary studies. This survey also provides a historical perspective of the evolution of the field and the role that humans can play in an increasingly automated SLR process.}




\maketitle


\section{Introduction}\label{sec:intro}

Artificial intelligence has come to alleviate people from tasks they repeatedly do at work but require some human abilities to success. Scientists are not an exception, and also demand powerful computational techniques to accelerate their results. In this sense, starting a new research often involves an in-depth analysis of related scientific literature in order to understand the context and find relevant works addressing the same or a similar problem. Besides, searching, screening and extracting key information from an extensive collection of papers is a time-consuming task that, doing without experience or clear guidelines, can lead to missing important contributions. Potential biases and errors can be mitigated by providing a rigorous methodology for literature search and analysis~\cite{Booth2016book}. A systematic literature review (SLR) is a secondary study that follows a well-established methodology to find relevant papers, extract information from them and properly present their key findings~\cite{Kitchenham2007}. The literature review is expected to provide a complete overview of a research topic, often providing a historical perspective which allows identifying trends and open issues. Literature reviews have become an important piece of work in many scientific disciplines, such as medicine --the area with the largest number of reviews published (13,510)-- and computing (6,342).\footnote{Source: Results of searching “Systematic literature review” on Scopus by February 1st, 2022.}

Conducting a literature review is known to be costly in time, specially if the authors cover a broad field. To support the SLR process, several tools have been created in the last years for different purposes~\cite{Marshall2013}. Among other features, SLR tools can import literature search results from electronic databases, mark them as relevant based on the inclusion criteria or provide visual assistance to analyse meta-information from authors and citations. Going one step further, automating the SLR process is gaining attention as an application domain in computing research~\cite{VanDinter2021}, mostly proposing methods that semi-automatically build search strings or retrieve papers from scientific databases. The use of automated approaches has proven to save time and resources when it comes to select relevant papers~\cite{Chapman2010} or sketch the report of findings~\cite{TorresTorres2017}. Nevertheless, some authors still suggest that their practical use is limited due to the required learning curve, and the lack of studies evaluating their benefits~\cite{Altena2019}.

In this paper, we focus on the automation of the SLR tasks using artificial intelligence (AI) as the main driver, seeking to augment the capabilities of automated methods and tools with additional knowledge and recommendations. The first use of AI techniques for automating SLR tasks dates back from 2006~\cite{Cohen2006}, when a neural network was proposed to automatically select primary studies based on information extracted via text mining. Following this idea, other authors have explored other text mining strategies~\cite{OMaraEves2015,Stansfield2017} and, more recently, machine learning (ML) and natural language processing (NLP)~\cite{VanDinter2021}. The possibilities that AI brings to the analysis of scientific literature are wide considering all the repetitive tasks that the SLR methodology entails. However, the role that humans play in the process should not be diminished, since they have an holistic view of the process that current AI techniques still lack.

The application of AI techniques to automate the SLR process is still a young discipline that is expected to continue growing in the next years. The increasing interest suggests that it is a good moment to analyse the AI techniques currently proposed to address the different SLR tasks, with special emphasis on their purpose, inputs and outputs, and human intervention, if any. Some of the secondary studies published so far in the area have already included AI techniques in their analysis of methods and tools for supporting SLR tasks. However, these studies either have been approached from a more general perspective, focusing on any kind of automation ---not necessarily focused on AI---~\cite{Marshall2013,VanDinter2021}, or are specialised in a particular AI technique (e.g. ML)~\cite{Marshall2019} or SLR task (e.g. paper selection)~\cite{OMaraEves2015, Olorisade2016}. Furthermore, these studies may lack an in-depth explanation of the AI concepts and techniques applicable to the whole SLR process. Therefore, this paper presents a complete survey of the area, while also seeks to deepen on the role that humans play in an semi-automatic SLR process, a perspective not considered by any previous literature review. With these goals in mind, we analyse the current state of AI-based SLR automation guided by the following research questions (RQs):

\begin{itemize}
\item RQ1. Which phases of the SLR process have been automated using AI?
\item RQ2. Which are the AI techniques supporting the automation of SLR tasks?
\item RQ3. To what extend is the human involved in SLR automation with AI?
\end{itemize}

To respond to these RQs, we conduct a systematic literature search as part of our survey. We identify 34 primary studies from more than 9,000 references retrieved from both automatic and manual search.\footnote{The search was completed up to the 30th of June of 2021.}
An analysis of these works is carried out to understand the purpose of using AI for solving a specific task. Then, we focus on the characteristics of the proposed methods, including their inputs, outputs and algorithmic choices. We also collect information on how the approach is experimentally evaluated, including the performance metrics and corpus of papers used for comparison. From our analysis, we found that some tasks are far more studied than others, and that some ML techniques proposed in the early stages are still used. However, we also discover some recent works exploring new ML approaches in which the human can be more involved. The discussion of our findings to answer each RQ has served us to identify some open issues and challenges related to unsupported tasks, additional AI techniques not still considered, and experimental reproducibility.


\section{Background}\label{sec:background}

A systematic literature review is a secondary study that rigorously unifies and analyses scientific literature in order to synthesise current knowledge, critically discuss existing proposals and identify trends. A SLR follows a well-established methodology to conduct evidence-based research~\cite{Kitchenham2007}, including the definition of research questions (RQs) and a replicable procedure to find relevant papers, a.k.a. primary studies, from which information will be extracted.

Conducting a SLR reports benefits to both its authors and the target research community. For authors, the SLR represents an opportunity to study a topic in depth, what is particularly recommended for graduate students~\cite{Felizardo2020}. For readers, SLRs provide a comprehensible and up-to-date overview of their field of interest, usually becoming a reference work to identify key studies and discover the latest advances. SLRs are known to have some drawbacks too, such as the long time needed to complete it or difficulties to evaluate the quality of primary studies~\cite{Kitchenham2013}. Recently, common threats related to SLR replicability have been analysed~\cite{Jacob2020}, pointing out problems that arise due to the lack of a clear methodology. The methodology proposed by Kitchenham and Charters~\cite{Kitchenham2007} divides the SLR process into the following phases:

\begin{enumerate}
    \item \emph{Planning phase}. The need for a SLR in the research area is motivated, thus guaranteeing that it will contribute to fill a gap and spread knowledge. Research questions are formulated to set the scope of the SLR and guide its development. They can follow predefined structures, e.g. PICO (Population, Intervention, Comparison and Outcome) or SPICE (Setting, Perspective, Intervention, Comparison and Evaluation)~\cite{Davies2011}. During this phase, a review protocol is prepared with a detailed strategy for all phases of the review. The protocol includes the search procedure and its sources, e.g. scientific databases and libraries; the definition of inclusion and exclusion criteria to select papers; and guidelines for data extraction and quality evaluation.
    
    \item \emph{Conducting phase}. Automatic searchers in databases and digital libraries are executed with search strings derived from the RQs or built with some supporting method~\cite{Mergel2015}. It is worth considering other sources too, such as grey literature and snowballing~\cite{Lefebvre2008}. The former consists in including sources like theses, dissertations, presentations and others that are not part of formal or commercial publications. Snowballing is a manual method where new literature is obtained by looking at references and citations in papers previously found. This helps access a more comprehensive collection of information on the topic. After the search, relevant studies have to be identified from the retrieved results, a process that includes duplication removal, identification of candidates ---usually based on title and abstract---, and the application of exclusion and inclusion criteria. These criteria specify the quality requirements that each paper must satisfy in order to be considered in scope~\cite{Booth2016}. The primary studies are then analysed to extract information. Summary statistics can be obtained to synthesise and visualise the collected data.

    \item \emph{Reporting phase}. This phase mostly refers to the writing process, including mechanisms to evaluate the completeness and quality of the final report. The authors should decide how the information is presented and discussed, and determine whether the review report is ready for publication. Guidelines in form of check-lists have been proposed to assess that the SLR report contains the essential information~\cite{Liberati2009}.
\end{enumerate}


\section{Methodology}\label{sec:methodology}

\begin{figure}[ht]
\centering
\includegraphics[width=0.8\textwidth]{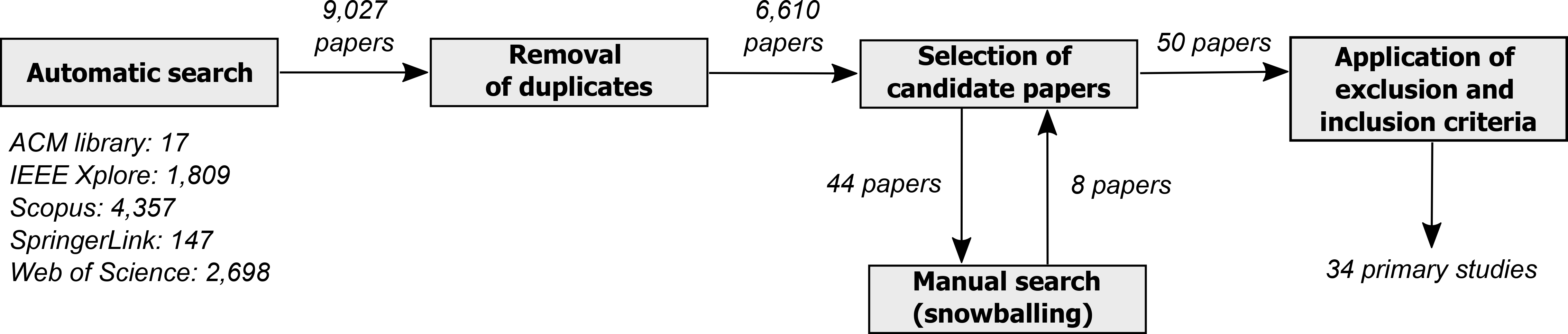}
\caption{Steps for searching and selecting relevant papers in AI-based SLR automation.}
\label{fig:methodology}
\end{figure}

Figure~\ref{fig:methodology} shows the methodological steps followed to retrieve papers and extract information from them~\cite{Kitchenham2007,Kitchenham2004}. Next, each step is explained in detail.

\subsection{Search strategy}\label{subsec:search}

The search strategy is comprised of both automatic and manual search. For automatic search, the following sources are queried: ACM Library, IEEE Xplore, Scopus, SpringerLink and Web of Science. The search string defined to retrieve papers is composed of multiple terms that combine keywords related to systematic reviews and words referring to automation. We choose general terms related to automation instead of a list of specific AI techniques for two reasons: 1) the list might bias the results to particular techniques, preventing less common approaches to appear in the results; and 2) a fully detailed list of techniques would result in large and complex search strings, which are difficult to manage by databases. Figure~\ref{fig:search-string} shows the resulting search string, which was conveniently adapted to each data source when needed. The fields considered for the search are title, keywords and abstract.

\begin{figure}[ht]
    \centering
    \begin{mdframed}
    \small\texttt{(((("Systematic Literature Reviews") OR ("Systematic Literature Review") OR ("Literature Review") OR ("Literature Reviews") OR ("SLR"))
    AND ("Automation" OR "Automated" OR "Automatic" OR "Automate"))
    OR ("Automated review"))} 
    \end{mdframed}
    \caption{Search string defined for retrieving papers related to SLR automation.}
    \label{fig:search-string}
\end{figure}

\begin{figure}[ht]
    \centering
    \includegraphics[width=0.8\textwidth]{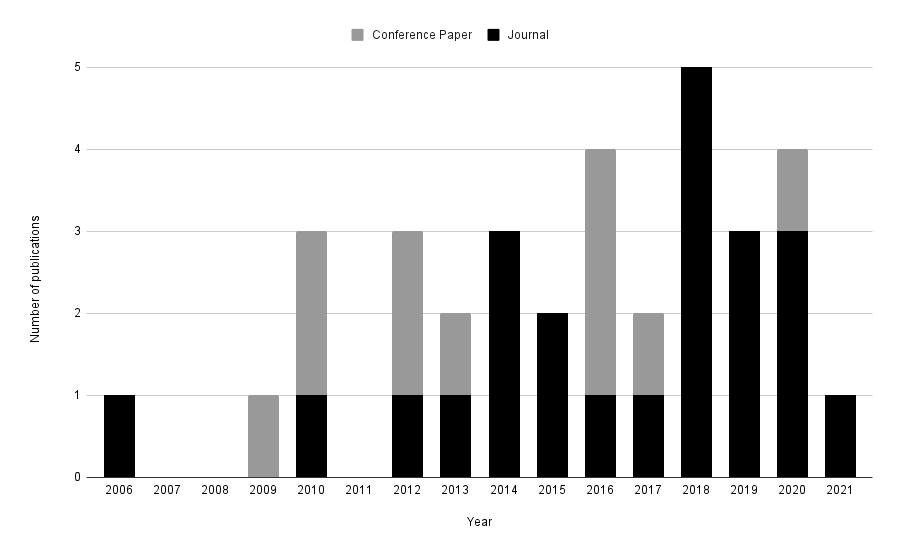}
    \caption{Distribution of primary studies per year.}
    \label{fig:primaryStudiesDistribution}
\end{figure}

After the execution of the search queries, 9,027 references are returned. Figure~\ref{fig:methodology} shows the number of papers retrieved from each source. From this set, 2,417 references are duplicates and, therefore, excluded from the total count. Then, a manual inspection of title and abstract is carried out to obtain a list of 44 candidate papers. Based on this list, manual search is performed via backwards snowballing. From the 8 papers initially found, 6 are added to the final list of candidate papers after reading their title and abstract. 

The 50 candidate papers are further analysed to confirm that they are within scope. With this aim, exclusion and inclusion criteria are established. Excluded papers correspond to manuscripts not written in English, those whose full content cannot be reached, or publications without evidence of a peer review process. Inclusion criteria specify restrictions applied to the content of the paper. To be considered for this survey, the paper should be focused on the automation of one or more phases of a SLR, and explicitly mention the application of some AI-based approach. This general criterion is decomposed into a number of mutually exclusive options: 1) the paper describes a new algorithm, tool or technique supporting the automation or semi-automation of a SLR; 2) the paper analyses the importance of the automation of a SLR and provides a retrospective of the state-of-art in this field; 3) the paper reports a summary of tools that are related to one or more phases of a SLR.

After applying these criteria, 34 papers are finally selected as primary studies. Figure~\ref{fig:primaryStudiesDistribution} shows their distribution along the years, divided into conference (32\%) and journal papers (68\%). The first study appeared in 2006, and it is not until 2009 that other proposals were published. After that, the number of papers per year remains more constant, without a clear predominance of conference or journals. However, it is noticeable that 57\% of the total journals papers (13) have been published in the last five years.

\subsection{Data extraction} \label{subsec:extraction}

Once all primary studies are identified, they are thoroughly analysed to gather information using a data extraction form~\cite{Kitchenham2007}. Each paper is revised by one author, a second reviewer being involved in case of any doubt. The data extraction form includes meta-information, e.g. authors and their affiliation, type of study and publication year, and categories to characterise the AI approach. More specifically, the content of each paper is summarised according to:

\begin{itemize}
  \item \emph{SLR phase and task}. The paper is classified according to the SLR phase(s) that it automates, detailing the specific step(s) in that phase.
  \item \emph{AI area and technique}. The paper is assigned to one or more AI areas, including a short description of the algorithm or method used.
  We also annotate if the human is somehow involved in the process.
  \item \emph{Experimental framework}. The type of primary study is identified among empirical, theoretical, application or review. For empirical studies, we collect the data corpus and the performance evaluation metrics used for evaluation.
  \item \emph{Reproducibility}. We revise if algorithms, datasets and tools included in the paper are publicly available. To do this, we check any website or repository mentioned as additional material to confirm that the content is reachable.
\end{itemize}


\section{AI techniques for SLR automation}\label{sec:survey-ai}

This section presents the AI techniques organised by SLR phase, namely planning (Section~\ref{subsec:survey-planning}), conducting (Section~\ref{subsec:survey-conducting}) and reporting (Section~\ref{subsec:survey-reporting}).

\subsection{AI techniques for the planning phase}\label{subsec:survey-planning}

At the beginning of the planning phase, it is recommended to perform a preliminary analysis of the scope and magnitude of the SLR~\cite{Stansfield2013}. In the context of health research, ``scoping'' reviews are a way to quickly identify research themes, for which papers need to be catalogued in order to obtain a ``map'' of the research topic. Due to its descriptive nature, unsupervised learning is suitable because it does not need data labels, i.e. predefined research topics in this case. In particular, clustering becomes a relevant approach here, as it is able to identify groups of entities like papers sharing characteristics. Lingo3G\footnote{\url{https://carrotsearch.com/lingo3g/} (Last accessed: February 14, 2023)} is a document clustering algorithm that has been used to group similar papers based on their title and abstract~\cite{Stansfield2013}. It allows papers to be associated to more than one cluster, and can also generate hierarchical clusters, thus providing a more refined topic classification. After clustering, the reviewer can map clusters to concepts. The method was evaluated using the results of previous ``scoping'' reviews from a health institution, comparing the topics automatically generated by clustering with those assigned by manual review.

Although the review process itself should be analysed during the whole SLR development, decisions about the available resources and task prioritisation should be taken during the planning stage. Process mining has been studied as a potential approach to understand the required effort and usual organisation of SLR activities~\cite{Pham2018}. Process mining encompasses, among other methods, a number of data mining techniques that analyse business processes by means of log events. Its main goal is the identification of trends and patterns with the aim of generating knowledge and increasing the efficiency of the business process. The method proposed by Pham et al.~\cite{Pham2018} analyses event logs produced by 12 manual SLR processes simulated by a multidisciplinary team. Logs represent the input to the process mining method, which is able to extract information about task assignment, timelines and effort measured in person-hour. More specifically, a heuristic mining algorithm analyses the frequency of events to determine the most relevant activities (e.g. searching papers, selecting them or reporting findings) and how they are temporarily distributed. To do it, a dependency graph is built to discover sequence patterns between the SLR tasks, e.g. whether a task is usually followed by another. Also, a fuzzy mining algorithm is executed to abstract different review models (how people conduct SLRs) by excluding less relevant activities or their characteristics (time spent, people involved, etc.). The algorithm uses two metrics, significance and correlation, to decide which events and relationships between them should be highlighted, aggregated or removed to simplify the process model.

\subsection{AI techniques for the conducting phase}\label{subsec:survey-conducting}

This phase has attracted great attention from the AI perspective, with 59\% of the primary studies related to its tasks. The selection of primary studies stands out as the most frequently supported task, with a total of 18 papers. ML is the most widely used branch of AI at this phase, often combined with NLP and text mining. Therefore, we first describe how paper selection is addressed from the ML perspective. Then, we focus on those tasks within the conducting phase that have been automatised with other different AI techniques.

The automatic selection of primary studies using ML requires two main steps: 1) the extraction of features to characterise the papers and 2) training a classifier to discern between those papers to be included and those to be excluded from the SLR. Feature extraction for paper selection often requires creating a list of topics or keywords from the title and abstract. NLP and text mining are applied to computationally handle and process such textual information. NLP provides efficient mechanisms for information retrieval and extraction from pieces of text so that they can be processed by a machine. NLP involves a series of steps to process and synthesise the data, such as word tokenisation, removal of stop words, and stemming. Text mining, which combines NLP steps with data mining methods, allows processing and analysing large fragments of text. Text mining is particularly relevant for inferring non-explicit knowledge and dealing with semantic aspects. In the second step, the list of candidate papers is processed by the learning algorithm based on their features, so that a decision is made about the relevance of the paper with respect to the SLR topic. In this case, three ML paradigms have been considered: supervised learning, active learning and reinforcement learning. In supervised learning, a labelled dataset is required to train the decision model. Active learning does not assume availability of labelled data, but considers that labels can be obtained at a certain cost. Reinforcement learning evaluates the rewards obtained when taking decision over the data.

\paragraph{\textbf{Supervised learning techniques for paper selection}}
Supervised methods have been extensively explored for paper selection, using existing SLRs to create labelled datasets to train from. The pioneering work combines text mining with neural networks~\cite{Cohen2006}. More specifically, the voting perceptron algorithm is used to train a classifier able to discern between relevant and non-relevant papers. The decision is based on a bag-of-words (BoW) representation of the papers, which is obtained from title and abstract via text mining using the Porter stemming algorithm and removing stop words. This work is also important because of the definition of the WSS\%95 evaluation metric, which has become a reference in many later studies. These authors use the same BoW representation in a subsequent study~\cite{Cohen2009}, which applies a fast implementation of support vector machine (SVM) called SVM$^{light}$. They also propose a novel way to train the model with a combination of topic-specific and non-topic-specific papers. By ``topic'' they refer to the research area for which the SLR is conducted, whereas non-topic papers are not strictly related to the field under study but to a close discipline. Such non-topic papers could be useful when the SLR covers a new research field with few publications yet. As the authors report, topic-specific classification can be biased and very few papers were deemed as relevant. In contrast, enlarging the training data with no-topic papers increased the performance of the method. In another study, SVM$^{light}$ is trained with 19 systematic reviews of different topics conducted in a medical institution~\cite{Kim2014}. Each paper is identified as included, excluded due to general criteria like the type of paper or publication source, or excluded due to topic-specific criteria. To characterise each paper, the authors combine the publication type with words extracted from title, abstract and indexing terms.

The performance of SVM and logistic regression (LR) with different set of features have been compared against human screening~\cite{Bannach-Brown2019}. A BoW approach is used to build the features for the SVM classifier, using the TF-IDF (Term Frequency-Inverse Document Frequency) metric to weight the importance of each word. As for LR, BoW features are combined with 300 topics extracted by a topic modelling algorithm (Latent Dirichlet Allocation, LDA). The authors study the performance of each method and the discrepancies between machine predictions and human decisions. Thomas et al.~\cite{Thomas2021} also consider a BoW approach, using title and abstract, to build an ensemble classifier. More specifically, the ensemble is comprised of two SVM models. The first SVM is trained with terms having one, two or three words in order to preserve some semantic. The second SVM only takes one-word terms into account and applies an oversampling method to improve the classification rate of the minority class. To put both SVM scores together into the ensemble, a logistic regression model known as Platt scaling is applied. This scaling generates an output in the interval [0,1], which represents  the probability that a paper is selected.

Naive Bayes (NB) classification is another supervised approach that has been studied for automating paper selection. FCNB/WE (Factorized Complement Naïve Bayes/Weight Engineering) combines a modified version of Complement NB (CNB) with feature engineering to assign different weights to the features. CNB amends the Multinomial NB (MNB) algorithm to use word count normalisation~\cite{Matwin2010}. A comparison against the algorithm proposed by Cohen et al.~\cite{Cohen2006} using the same corpus of papers is included to assess the improvement achieved by their proposal. CNB has been trained under two additional different methodologies to deal with the imbalanced training set of candidate papers~\cite{Frunza2010}. First, the authors use a human-annotated training corpus for which three representations are compared: 1) BoW like in \cite{Cohen2006}, 2) a more specialised collection of terms from a medical knowledge repository, and 3) the combination of both. Only abstracts are considered to classify papers, simulating an early step of candidate identification. The second approach, referred as per-question classification, requires building a classifier for each inclusion criterion. Different voting aggregation methods are studied to finally decide whether the candidate paper is selected or not. Definitely, SVM and NB are the supervised techniques most frequently applied, even though other classification algorithms have also been employed. Garc\'ia Adeva et al.~\cite{Adeva2014} combine the use of seven feature selection methods and four classification techniques. Feature selection is applied to keep only a proportion of the most relevant terms, which are measured using popular text mining metrics like term frequency (TF), document frequency (DF) and inverse document frequency (IDF). As for classification, the authors compare NB, SVM, k-nearest neighbours (kNN) and Rocchio. The experimentation suggests that SVM outperforms the other algorithms when the papers are characterised by their title or by a combination of title and abstract. When only abstracts are considered, Rocchio and NB show better performance. Almeida et al.~\cite{Almeida2016} present another study comparing several classifiers and feature sets, which is specialised for biomedical literature. The papers are represented as BoW taken from abstract and title, alone or in combination with a specific list of biomedical terms. They select a subset of words using two metrics: IDF and Odds ratio. The authors compare NB, SVM and a logistic model tree (LMT), which builds a decision tree (DT) using logistic regression models on the nodes. The best results were obtained by LMT over the combination of BoW with biomedical terms.

\paragraph{\textbf{Active learning techniques for paper selection}}
All the above methods work under a supervised strategy.  Note that in these cases, the corpus of candidate papers could be comprised of thousands of irrelevant papers retrieved by automatic search if the search string is too generic or not sufficiently refined. Active learning has appeared as a relevant paradigm for paper selection, since it is founded on the idea that labelling is a costly process that can be only partially done by querying an external oracle during the learning process. The classification can be performed by usual techniques for supervised learning, SVM being the preferred one for paper selection. Based on this idea, Abstrackr applies SVM under an active learning approach where the oracle is the human reviewer~\cite{Wallace2012}. Implemented as a web tool, Abstrackr shows the title, keywords and abstract of a paper to be labelled as relevant, irrelevant or borderline. Reviewers are asked to highlight those terms that support their decision, which will be exploited then for learning by the SVM classifier.

The labels annotated by a human reviewer can be propagated to similar unlabelled papers following different strategies~\cite{Kontonatsios2017}. One possibility is that the label assigned by the reviewer is propagated to neighbouring unlabelled papers using the cosine distance between the paper representations: BoW or a low-dimensional representation obtained by a technique similar to principal component analysis. The underlying classifier, SVM, predicts the label of the remaining papers together with a certainty level. In each new cycle, the reviewer is asked to provide new labels for a sample of either the less or the more uncertain predictions. \hbox{FASTREAD}~\cite{Yu2018} is a conceptual active learning approach also using SVM as the underlying classifier, which can be ``instantiated'' into 32 different learning models depending on: 1) when to start training, 2) which document to query next, 3) when to stop training, and 4) how to balance the training data. The 4,000 terms from title and abstract with highest TF-IDF score become the features for learning. The authors are particularly interested in analysing the ability of these methods to exclude irrelevant works, showing that a specific configuration of their abstract method leads to better performance than state-of-the-art algorithms. Build upon these findings, a later work presents FAST$^2$, an improved active learner~\cite{Yu2019}. FAST$^2$ includes a new strategy to identify the first relevant paper using domain knowledge, a LR-based estimation to decide when learning should stop, and a method to revise disagreements in paper labelling between the learner and the human. 

\paragraph{\textbf{Other methods to support paper selection}}
As suggested above, the selection of primary studies is strongly related with the quality of the search, so the first task could benefit from an automatic definition of search strings too~\cite{Ros2017}. The method starts from an initial set of accepted papers, whose title, abstract and keywords are used to infer the search strings by means of a DT (ID3 algorithm). Automatic search is then executed to collect candidate papers, which will undergo the ML-based paper selection. First, a BoW representation, extracted from title, abstract and keywords, is combined with a list of topics discovered by LDA to build the features. Since the authors argue that paper selection should be interactive and iterative, they propose the use of semi-supervised learning approaches: active learning (AL) and reinforcement learning (RL). The former will show the reviewer those papers with the highest probability of being primary studies, or those for which the classifier is more uncertain. The latter combines both ideas (probability and uncertainty) to explore papers that are not necessarily the most relevant ones as a way to avoid local optima. SVM and LR are internally used as classifiers for AL and RL, respectively. The authors also include greedy approaches of SVM and LR that automatically select the paper with highest probability.

Some other AI-based techniques have been proposed to assist in the process of paper selection, but they are not directly intended to automatically select the set of primary studies. Rather, the pool of candidate papers is inspected with additional information in order to evaluate their quality. In a first study, text mining and interactive visualisation techniques are combined~\cite{Felizardo2012}. In visual text mining, visualisation techniques are incorporated to show relations between documents and help inspecting textual data~\cite{Alencar2012}. These techniques are used to build a ``document map'' showing the relationships among candidate papers based on content similarity. Content similarity is calculated as the cosine distance between papers, represented as a BoW vector. The extracted words are weighted using the TF-IDF metric. Clustering using the k-means algorithm is applied over the map, whose results should be later analysed by the reviewer using additional information. For instance, a citation map showing co-citation relationships extracted from bibtex files can be used to decide the quality of the paper. The visual analysis is supported by Revis, a tool for document map creation, which was extended to incorporate citation maps. In a subsequent study, the authors propose the score citation automatic selection (SCAS) strategy, which again combines paper content and citation information to select candidate papers~\cite{Octaviano2015}. Two tools support their method: StArt that provides a classification score based on the frequency of appearance of the search string in title, abstract and keywords; and Revis, for the analysis of cross-references among research papers. SCAS takes two inputs, the StArt score and whether the candidate paper is cited or not, to train a DT (J48 algorithm). The tree classifies the papers into four classes (included, excluded or two categories of ``to be reviewed''), also allowing to identify the cut-off point of the Start score that separates included papers from excluded ones. Labels are obtained from manual selection using three SLRs as case studies. Thirdly, the work by Langlois et al.~\cite{Langlois2018} automatically classifies papers into empirical and non-empirical studies. The former are considered as relevant, while the latter are discarded. kNN, NB, SVM, DT and ensembles (bagging and boosting strategies for DT) are applied as classification techniques. In this case, the authors first build the classification models with words extracted from title, abstract and a thesaurus of medicine terms. Then, they analyse the classification performance under different ratios of full-text availability, concluding that adding words from full texts slightly improved the obtained results. 

\paragraph{\textbf{AI techniques for data extraction and summarisation}}
Finally, a few AI techniques are focused on the data extraction task with the purpose of supporting knowledge representation. In this sense, ontologies are the main mechanism to capture real-world concepts and their semantic relationships. Ontology-based systems use a representation language, e.g. first-order logic or fuzzy logic, to encode such knowledge, which is combined with automatic reasoning techniques to make inferences. In the context of automated data extraction, the SLROnt ontology defines the concepts that appear in two key elements of a SLR: the review protocol and the set of primary studies~\cite{Sun2012}. The method is focused on automatic reasoning about primary studies, using abstract information to describe their most important characteristics. Such a description is based on the usual categories of structured abstracts (background, objective, method, results and conclusion). Similarly, the use of ontologies with information extracted from abstracts has been proposed as a means of providing a short description of biomedical papers~\cite{Erekhinskaya2016}. A semantic representation of each paper is then derived, mapping words to concepts from three medical ontologies and setting predefined relationships among them. The paper description is generated from the semantic information by filling a PICO-based template. ML is applied for entity recognition during concept parsing, even though the details of the algorithm are not provided in the paper. 

Data extraction has been treated as a learning problem too, whose goal is to classify relevant sentences for summarising experimental results~\cite{Lucic2016}. In particular, this method identifies key sentences about medical treatment comparisons from full texts. SVM classifiers with linear and Gaussian kernel methods are trained with 100 sentences using words and concepts manually assigned. The method works under a multi-class approach, trying to identify the entities and treatment characteristics that appear in the comparison sentences. 

\subsection{AI techniques for the reporting phase}\label{subsec:survey-reporting}

This last phase of the SLR process has received little attention yet. Current AI approaches only support two tasks: writing the SLR report and its evaluation.

The automatic generation of content for the SLR report is a complex task not addressed until very recently. A summary of each selected primary study is a good starting point to write a SLR report. Teslyuk et al. envision a system combining NLP and deep neural networks able to generate such summaries~\cite{Teslyuk2020}. Deep learning is suitable here due to its ability to learn complex concepts from simple ones using layered architectures. The conceptual model takes a set of papers as input, for which up to five sentences located around citations are extracted using NLP. A pre-trained biomedical language representation model, called BioBERT, is responsible for encoding the sentences that will be transformed into summaries by means of a long short-term memory (LSTM) recurrent neural network. A LSTM efficiently processes sequences of data, e.g. text, allowing to keep and forget parts of the inferred information.

A way of evaluating the SLR report is to analyse whether the relevant aspects of the primary studies are well reflected in the report. To do this, Liu et al.~\cite{Liu2010} propose the use of NLP to generate automatic questions about the content of the papers. These questions address the subject of research, its aim and contributions, the method and datasets used, the results obtained and the strengths and limitations of the method. A name entity tagger, called LBJ, is the NLP technique applied for automatic question generation, together with phrase parsers and regular expressions. LBJ has a language model based on functions, constraints and an inference mechanism to support NLP tasks such as part-of-speech tagging, chunking and semantic labelling~\cite{Rizzolo2007}. In the primary study, LBJ automates the identification of author names in citations. Then, the method formulates questions about the sentence explaining the cited work.

\subsection{Previous analyses of the field and tool evaluations}\label{sec:tools}

During the literature search, we found works that cannot be classified in a particular phase. These works compare existing tools or analyse research literature related to the use of AI for SLR automation. They complement our analysis from different viewpoints and allow us to obtain a historical perspective.

A mapping study of tools to support SLRs in a computing field (software engineering) is based on the analysis of 14 papers~\cite{Marshall2013}. The authors found that text mining, including those that integrate visualisation techniques, are prevalent in the area (57\%). Extensions of Revis and the SLROnt ontology mentioned in Section~\ref{subsec:survey-conducting} appear in this study, as they were evaluated with corpus of papers related to software testing and cost estimation, respectively. The authors conclude that the analysed tools were at an early stage of development. Besides, experiments to assess their effectiveness were still very preliminary.

Tsafnat et al.~\cite{Tsafnat2014} provide an overview of SLR automation in the domain of evidence-based medicine. Focusing on AI-based tools, they only include Abstrackr (see Section~\ref{subsec:survey-conducting}) in their analysis. Other techniques, like ontologies, clustering, supervised classification and NLP are mentioned but as part of reference managers and specialised bio-medical systems without providing an in-depth analysis. In addition, the authors see great potential on the application of AI for: 1) automatic hypothesis generation, 2) improvements on inclusion criteria through reasoning, 3) duplication detection via NLP, 4) abstract screening combining ML and heuristics, and 5) better text analysis using NLP and optical character recognition for multi-language support. 

Two other secondary studies are focused on the analysis of ML techniques for the paper selection task~\cite{OMaraEves2015,Olorisade2016}. The former provides a retrospective of different approaches to analyse how they contribute to workload reduction and the challenges that their application entail. From their analysis, the achievable workload reduction greatly varies depending on the experiments (30-70\%). Among the identified problems, the authors highlight imbalanced data, i.e. the percentage of relevant studies is very low compared with the number of non-selected papers. They suggest that class weighting and undersampling are possible solutions to this problem. Focusing on the techniques, the authors conclude that active learning ensures higher recall. The second work presents a more detailed analysis of text mining techniques required for preprocessing as part of paper screening. The studied techniques are characterised in terms of the method used to extract features for learning, the type of classifier, performance measures for evaluation and corpus of papers. Feature representation is mostly based on term frequency (66\%), including works that use TF-IDF and other information gain metrics to weight words. They found 13 different classification algorithms, SVM and ensembles being the most widely applied.

A different perspective of the field is provided in two recent studies~\cite{Marshall2019,Beller2018}. On the one hand, Beller et al.~\cite{Beller2018} present the principles that should guide the development of automated methods for SLR, which were derived from an international meeting of members of the ICASR (International Collaboration for the Automation of Systematic Reviews) group. The desired principles include improvements in efficiency, coverage of multiple tasks, flexibility to use and combine methods, and better replicability promoting the use of open source resources, among others. On the other hand, Marshall et al.~\cite{Marshall2019} develop a practical guide for the use of ML methods to conduct SLR in the medicine domain. The study is conceived as an introduction for non-experts, discussing the scope of each tool, as well as their strengths and limitations. Therefore, they only analyse tools accessible in an online catalogue named SR Toolbox,\footnote{\url{http://systematicreviewtools.com/} (Last accessed: February 14, 2023)} omitting scientific literature unless a supporting tool is also available. 13 tools are analysed, classified depending on the SLR task: literature search, paper selection and data extraction. The authors suggest that most of these tools should be viewed as assistant tools, where the user plays a key role in validating the provided results. However, they also prevent about the usability of these tools, since most of them are still prototypes or research-oriented tools. Nonetheless, new tools have appeared and others have evolved in the last years. We provide an up-to-date analysis of SLR tools using AI in our supplementary material.


\section{Analysis of current trends}\label{sec:findings}

We discuss the state of the field in terms of SLR phases currently supported (RQ1), the selection of AI techniques (RQ2) and human intervention (RQ3).  

\subsection{SLR phases currently supported}\label{subsec:findings-rq1}

Focusing on RQ1, our literature analysis indicates that all phases of the SLR process have been covered by at least one primary study, but that the conducting phase stands out as the most studied by far due to the strong interest on the automatic selection of primary studies. This prevalence is in line with the conclusions drawn by the most recent review on SLR automation~\cite{VanDinter2021}. In contrast, this review also concluded that no study, either using AI or not, supports the planning and reporting phases, although there are some primary studies applying AI techniques used in these phases, as explained in Section~\ref{subsec:survey-planning} and Section~\ref{subsec:survey-reporting}. The effort required during the selection of primary studies might explain well the high number of AI proposals to automate it. Indeed, several studies have measured the time spent on manual and semi-automatic selection, suggesting that AI-based methods can reduce screening burden up to 60\%~\cite{Tsou2020} and represent time savings of more than 80 hours~\cite{Gates2020-1}. However, only a couple of tools supporting paper selection, Abstrackr and EPPI-Reviewer, seem to be relatively popular in the medicine domain. The fact that most of the proposed methods are not available as tools or integrated in other systems like reference managers seems to be hampering its use in practice. This is also applicable to the rest of phases and tasks, since most of the surveyed publications only cover a very specific problem without giving complete support to the SLR process. According to our findings, only two papers address more than one task~\cite{Pham2018,Ros2017}. 

From a historical perspective, it is also interesting to note that the selection of primary studies continues to attract attention since the publication of the first paper~\cite{Cohen2006}. Five new methods have been proposed in the last four years~\cite{Bannach-Brown2019,Yu2019,Thomas2021,Yu2018,Langlois2018}, and supporting tools are subject of evaluations~\cite{Tsou2020,Marshall2019,Gates2020-1}. 

\subsection{Selection of AI techniques}\label{subsec:findings-rq2}

In response to RQ2, ML is the most frequent AI area, with contributions exploring different learning paradigms: supervised and active learning for classification and, less often, unsupervised learning for clustering. Active learning has become the reference approach for paper selection~\cite{Yu2019,Kontonatsios2017,Yu2018,Ros2017}. With this approach, the cost of labelling is explicitly modelled, not assuming endlessly availability of previously labelled training data. Another recurrent characteristic is imbalance during the paper selection task, for which authors have selected algorithms specifically designed for problems with imbalanced class distribution~\cite{Frunza2010}, or have incorporated some data balancing technique~\cite{Thomas2021,Yu2018}.

Focusing on ML algorithms, SVM is frequently adopted for classification (13 out of 17 papers), either under supervised or active learning approaches. SVM is known to be highly effective to cope with high dimensional feature spaces~\cite{Cervantes2020}, as is the case of the paper selection problem using a BoW feature representation. The rest of classifiers explored for paper selection are NB (5), DT (3), LR (2) and neural networks (2). Nevertheless, the number of papers is rather low to draw conclusions about why a particular algorithm was chosen.

Since most of the primary studies are focused on the paper selection problem, we further analyse the characteristics of the methods in terms of required inputs, types of outputs and availability of paper corpora for training. Table~\ref{table:inputTable} summarises this information for the 11 papers focused on paper selection. Text mining is the usual approach to extract representative words from the candidate papers, which are later used to build the features for learning. In general, words are obtained from the title and abstract, and less often from the keywords too. Inspecting only these parts of candidate papers is the standard procedure during manual screening~\cite{Kitchenham2007}, and the most common approach in SLR automation~\cite{VanDinter2021}. However, text mining techniques are powerful enough to manage large pieces of text, so AI methods could increase the amount and quality of the information used. This would allow including more details about the paper content that might not appear in the header section, i.e. title, abstract or keywords, but at the expense of many more words to be processed. To reduce the dimension of the feature space, many authors rely on scoring methods, such as TF-IDF, to weight the words and keep only the most representative terms. Another alternative is the application of the LDA algorithm, which allows setting a predefined number of high-level topics to be extracted. In terms of tools, Abstrackr is more flexible in this sense, because it lets researchers interactively highlight the relevant and irrelevant words at their convenience~\cite{Wallace2012}.

\begin{table}[h]
\begin{center}
\caption{Inputs, outputs and data corpus used by primary studies for paper selection.}\label{table:inputTable}%
\begin{tabular}{@{}p{3.5cm} p{4cm} p{2cm} l@{}}
\toprule
\textbf{Inputs} &	\textbf{Outputs} &	\textbf{Corpus (domain)} & \textbf{Ref.} \\
\midrule
 Topics                      & Ranking (0-1) of papers       & 1 SLR (M)*                &\cite{Bannach-Brown2019}\\
 Title, abstract,            & Paper classification in four  & 3 SLR (C)                &\cite{Octaviano2015}\\ 
 keywords and cites          & categories and citation map   &                           & \\ 
 Title and abstract          & Binary (selected papers)      & Same as \cite{Yu2018}  &\cite{Yu2019}\\ 
 Title and abstract          & Binary (selected papers)      & 1 SLR (M)                 & \cite{Cohen2006}\\
 Title and abstract          & Binary (selected papers)      & 1 SLR (M)                 & \cite{Cohen2009}\\
 Title, abstract, keywords   & Binary (selected papers)      & 1 SLR (M)                 & \cite{Kim2014}\\
 and publication type        &                               &                           & \\ 
 Title and abstract          & Binary (selected papers)      & 1 SLR (M)                 &\cite{Thomas2021}\\
 Title, abstract, keywords   & Binary (selected papers)      & Same as \cite{Cohen2006}  & \cite{Matwin2010}\\
 and publication type        &                               &          &  \\
 Title and abstract          & Binary (selected papers)      & 1 SLR (M)                 & \cite{Frunza2010}\\ 
 Title and/or abstract       & Binary (selected papers)      & 1 SLR (M)                 &\cite{Adeva2014}\\  
 Title, abstract             & Binary (selected papers)      & 1 SLR (M)                 &\cite{Almeida2016}\\  
 and specific terms          &                               &                           & \\ 
 Title, abstract, keywords   & Relevant, irrelevant          & 1 SLR (M)                 & \cite{Wallace2012}\\
 and user-defined words      & borderline papers             &                           & \\ 
 
 Title and abstract          & Binary (selected papers)      & 6 SLR (M)                 &\cite{Kontonatsios2017}\\
 Title and abstract          & Binary (selected papers)      & 4 SLR (C)*               &\cite{Yu2018}\\ 
 Title, abstract, keywords,  & Ranked probability of a paper & 1 SLR (N/A)               & \cite{Ros2017}\\
 topics and metadata         & to be selected in the SLR     &                           & \\ 
 Title and abstract          & Document and citation map     & 4 SLR (C)                &\cite{Felizardo2012}\\ 
 Title, abstract             & Binary (selected papers)      & 5 SLR (M)                 &\cite{Langlois2018}\\
 and specific terms          &                               &                           & \\
\botrule
\end{tabular}
\footnotetext{Acronyms: M=medicine, C=computing, N/A= Not available, *= Available online.}
\end{center}
\end{table}

Guidelines for SLR often refer to criteria based on meta-information or quality for defining the selection strategy in the review protocol. Language, extension or type of publication are exclusion criteria that can greatly reduce the number of candidate papers to be inspected. Despite this, very few works include features beyond the paper content. Only two methods complement word processing with other kind of information, citations and cross-references~\cite{Felizardo2012,Octaviano2015}. In both cases, visualisation mechanisms and clustering methods are developed to build assistant tools that facilitate the analysis. The rest of algorithms perform classification in one step, i.e. a binary decision of whether the paper should be selected or not. Breaking with this idea, a few methods~\cite{Ros2017,Bannach-Brown2019} propose that the output should be a ranking, similarly to \hbox{Abstrackr}, where researchers can rate papers as relevant, irrelevant or borderline~\cite{Wallace2012}. Overall, most of the AI-based methods detect a reduced list of papers within scope, not really simulating a criteria-guided evaluation.

\subsection{Human intervention}\label{subsec:findings-rq3}

During the data extraction process, the need of human intervention was carefully observed in order to respond to RQ3. AI approaches were classified as fully automated (68\%) or semi-automated (32\%). The former case corresponds to those primary studies for which the human does not take part in the execution of the AI approach. This category includes supervised learning techniques and any other method requiring an input corpus of papers, even if it is previously created or annotated by a human. Hence, semi-automated approaches should mention explicitly that some kind of human intervention is required.

Abstrackr is an interactive tool whose classifier is trained based on the feedback provided by one or more reviewers. More specifically, they can perform two actions: 1) highlighting relevant and irrelevant words within the title and abstract; and 2) marking the paper as accepted, rejected or borderline. For borderline papers, reviewers also have to introduce the number of SLRs that they have conducted in the past as an indicator of their expertise. Then, borderline papers are shown to more experienced reviewers. Abstrackr is the AI-based tool that has been adopted by more independent researchers to evaluate its performance. In such studies, participants have been asked to use Abstrackr to reproduce the paper screening of real SLRs with the purpose of measuring the time saved and the precision of the final paper selection.

The rest of active learning methods mention humans as an oracle for providing labels, though the presented experiments do not involve actual participants. Kontonatsios et al.~\cite{Kontonatsios2017} use the label assigned to one paper by the human reviewer to tag other similar papers that remain unlabelled. The authors present two strategies to decide which papers should be shown to the human: 1) choose the more relevant papers according to the classifier, or 2) let the human classify those papers for which the classifier has less confidence in its prediction. In the experiments both approaches are automatically evaluated taking a percentage of labelled papers from a training set, showing that the classifier can achieve 92\% performance with only 5\% of labelled papers. Such a percentage seems manageable for a scenario of collaboration with a human.

Ros et al.~\cite{Ros2017} present a proof-of-concept in which the reviewer should validate papers suggested by the tool. The information displayed to the human includes the most relevant terms used by the classifier to make a decision, as well as information about how the paper was found, i.e. snowballing or automatic search. The papers to be validated are selected following two strategies: 1) picking papers close to the decision boundary built by the classifier, and 2) promoting papers predicted as positive by the classifier. The experimental validation is automatically performed by looking the manual labels assigned within a training set created from a SLR previously conducted by the authors. 

For their general FASTREAD method, Yu et al.~\cite{Yu2018} explore the same strategies as Ros et al.~\cite{Ros2017}. The authors discuss that it would be desirable to allow having multiple reviewers, assigning different sets of papers to each one. This idea represents a challenge since the ML algorithm would need to deal with potential human disagreements. This particular problem is addressed in a subsequent work~\cite{Yu2019}, but still focused on only one human reviewer. Here, FAST$^2$ analyses the class probability estimation each time the human oracle labels 50 new papers, and those papers on which the active learner and the human reviewer strongly disagree are marked to be rechecked. To test their strategy, the authors simulate inconsistencies in the human evaluation.

\section{Open issues and challenges}\label{sec:issues}

We have identified a number of open issues that lead to challenges: 

\noindent\textbf{\emph{One single task is predominant.}} Research into SLR automation with AI is strongly biased towards the conducting phase and, more specifically, the paper selection task. Although this task is time consuming, the application of AI to other tasks demands attention. Some initial works have appeared, but are less mature compared to the algorithms proposed for paper selection. We identify AI-driven writing tasks, e.g. formulating RQs, defining exclusion/inclusion criteria or reporting SLR results, as the main challenge in this direction.

\noindent\textbf{\emph{AI techniques are still to be explored.}} The spectrum of AI areas and techniques is wide, but some of them have not been applied to SLR automation yet. For instance, optimisation and search techniques have not been explored for any SLR task resolution. This type of techniques have been traditionally used to solve planning problems, thus we speculate that they could be applied during the first phase to prioritise resources, e.g. choose the best databases, or distribute work, e.g. assign papers to reviewers based on their skills. Compared to ML, knowledge representation and NLP appear less often and most of the proposals seem to be in an initial stage. Consequently, there is a lack of tools and frameworks to develop solutions based on these techniques. 

\noindent\textbf{\emph{Specialised algorithms can replace general purpose approaches.}} Focusing on ML for paper selection, SVM has become a reference algorithm, probably due to the choice of the high-dimensional BoW representation. It would be interesting to study the applicability of other algorithms under the same or other feature spaces. The need of approaches specifically designed for the paper selection problem and for other tasks in SLR automation, should be explored in-depth. Some challenges here are related to the combination of types of input information to enrich the process, as well as to obtaining more flexible outputs beyond selected/non-selected. For SLR tasks requiring text analysis, the methods must be retrained or adapted to learn from the specific vocabulary of the scientific discipline (medicine, computing, etc.) under review.

\noindent\textbf{\emph{More complete information can improve decision-making.}} As for the features, BoW representation of title and abstract clearly dominates. Content from different paper sections, as well as meta-information and citation analysis, may be considered as well. Nevertheless, strongly relying on paper content implies that the classifier can only use the ``vocabulary'' of the field to make decisions, missing those papers adopting a different or emerging terminology, or simply those covering new or disruptive topics. Therefore, the analysis of related research communities, including co-authorships and cross-references, could be necessary to identify emergent topics for which a standardised terminology has not been comprehensibly developed yet.

\noindent\textbf{\emph{More active human involvement can benefit AI.}} The level and nature of the cooperation between the human and the AI methods or tools is still limited. At the moment, the role of the human is mostly oriented towards providing some labels for paper selection under an active learning approach. The planning and writing phases, which clearly demand more human skills, could benefit from interactive AI. Involving humans in this process would also have other beneficial effects, such as adapting the results to their preferences.

\noindent\textbf{\emph{End-users of SLR automation are not necessarily AI experts.}} Most of the ML techniques considered so far ---e.g. SVM or neural networks--- are known as ``black-box'' techniques. The fact that SLRs are conducted by scientists from diverse disciplines, not necessarily experts in AI, poses the challenge of the lack of trust in automatic results. Interpretable models, such as rule-based systems or small decision trees, have been barely explored. Also, we envision that the potential of recent explainable methods would allow complementing the output of black-box AI solutions developed in this area.

\noindent\textbf{\emph{AI-based automation of SLRs can be scaled up.}} Most of current proposals have been validated in the field of medicine or computing, sometimes using domain-specific ontologies or concepts to build the feature set. Probably the hardest limitation here lies in the availability of benchmarks, since real SLRs are not always fully replicable. Even when the set of candidate, excluded and included papers is available, decisions made for their selection might not be explicitly linked to inclusion and exclusion criteria. Further progress should be made in extending the evaluation of AI methods to cover a wider variety of SLRs, as well as broadening the scope of topics.

\noindent \textbf{\emph{Performance comparison between different methods and fields.}} The performance of AI-based techniques for SLR automation has been studied for fields like medicine or computing. Applying one technique to solve the same SLR task in a different field may not be trivial due to the specific terminology or types of research papers of each field. Studying the applicability of techniques to different fields is necessary to determine how they should be adapted. It would also be useful to compare methods to find out to what extent their performance depends on the application field, or if there are methods that fit better than others to the specific characteristics of a given field of knowledge.

\noindent\textbf{\emph{Open science fosters the development of practicable methods.}} In terms of reproducibility, the availability of implementations and corpora is still rare. Some tools and algorithms were originally made public but they are not accessible any more. As interest in this area continues to grow, there is an increasing need to provide access to algorithms and to set common experimental frameworks that allow comparisons between proposals. This point is seemingly less challenging, but still requires considerable effort from the community to make artefacts not only accessible but also fully functional.

Finally, we provide suggestions based on our own experience when trying to use some of the reviewed methods to accelerate SLR tasks. In particular, we tested two paper selection tools (Abstrackr~\cite{Wallace2012} and FAST$^2$~\cite{Yu2019}) to replicate our own search for primary studies. FAST$^2$ was considerably more effective than Abstrackr, since we were able to find almost 95\% of the primary studies with less than 10\% of screened papers. In contrast, Abstrackr found only 10\% of the primary studies after screening the same number of papers (300). Despite some configuration issues due to the requested dataset format, we found these tools useful and intuitive. We suggest some improvements regarding the information shown to the reviewers, e.g. why a paper was selected, and how they can add information to improve the process, e.g. by adding key words at the beginning instead of iteratively. Even if some tool support is available, we consider that the success of an SLR still lies on researchers' shoulders in terms of methodological steps (clear review protocol, checkpoints for replication) and analytical capabilities (summary of papers and trends analysis).

\section{Conclusions}\label{sec:conclusions}

The application of artificial intelligence has shown to be effective in automating many tasks humans find costly and repetitive to do, as is the case of conducting literature reviews. Planning, conducting and reporting a SLR involve many individual tasks, so it not surprising to observe that not all of them have been automatised yet. Our findings reveal a clear interest in applying AI, specially ML, to support paper screening, a burden task aimed at identifying relevant works from thousands of candidate papers. Regarding other tasks, we can highlight the use of ontologies and NLP to deal with semantic information. Nevertheless, the number of studies in these areas are still far less abundant.

Future efforts should be devoted to provide support to the planning and reporting phases, whose tasks are more difficult to automate.
Advances in automatic writing would be expected in the near future because of the appearance of some conceptual approaches based on deep learning.

\backmatter

\bmhead{Supplementary information}

Detailed results of the literature search and an analysis of tools are available from: \url{https://www.uco.es/kdis/ai4slr/survey}.

\bmhead{Acknowledgments}
Grant PID2020-115832GB-I00 funded by MICIN/AEI/10.13039/501100011033. Andalusian Regional Government (postdoctoral grant DOC\_00944).

\bmhead{Competing interests}
The authors have no competing interests to declare.

\bmhead{Declarations}
Conceptualization: all authors; Methodology: all authors; Formal analysis and investigation: J. de la Torre-López, A. Ram\'irez; Writing - original draft preparation: J. de la Torre-López; Writing - review and editing: A. Ram\'irez, J.R. Romero; Funding acquisition and supervision: J.R Romero.


\bibliography{references}


\end{document}